\documentclass[twocolumn,showpacs,preprintnumbers,superscriptaddress,
amsmath,amssymb]{revtex4}

\usepackage{epsfig}
\usepackage{graphicx}
\usepackage{dcolumn}
\usepackage{bm}
\usepackage{times}
\usepackage{xcolor}

\begin{document}


\title{Social contagions with communication
  channel alternation on multiplex networks}

\author{Wei Wang}
\affiliation{Cybersecurity Research Institute,
Sichuan University, Chengdu 610065, China}
\affiliation{Big Data Research Center, University of Electronic Science
  and Technology of China, Chengdu 610054, China}

\author{Ming Tang}\email{tangminghan007@gmail.com}
\affiliation{School of Information Science and Technology,
East China Normal University, Shanghai, 200241, China}
\affiliation{Shanghai Key Laboratory of Multidimensional Information Processing, East China Normal University, Shanghai 200241, China}

\author{H. Eugene Stanley }
\affiliation{Center for Polymer Studies and Department of Physics,
Boston University, Boston, Massachusetts 02215, USA}

\author{Lidia A. Braunstein}
\affiliation{
Instituto de Investigaciones F\'{i}sicas de Mar del
Plata (IFIMAR)-Departamento de F\'{i}sica,
Facultad de Ciencias Exactas y Naturales,
Universidad Nacional de Mar del Plata-CONICET,
Funes 3350, (7600) Mar del Plata, Argentina}
\affiliation{Center for Polymer Studies and Department of Physics,
Boston University, Boston, Massachusetts 02215, USA}

\date{\today}

\begin{abstract}

\noindent
Internet communication channels, e.g., Facebook, Twitter, and email, are
multiplex networks that facilitate interaction and information-sharing
among individuals. During brief time periods users often use a single
communication channel, but then communication channel alteration (CCA)
occurs. This means that we must refine our understanding of the dynamics
of social contagions. We propose a non-Markovian behavior spreading
model in multiplex networks that takes into account the CCA mechanism,
and we develop a generalized edge-based compartmental method to describe
the spreading dynamics. Through extensive numerical simulations and
theoretical analyses we find that the time delays induced by CCA slow
the behavior spreading but do not affect the final adoption size. We
also find that the CCA suppresses behavior
spreading. On two coupled random regular networks, the
adoption size exhibits hybrid growth, i.e., it grows first continuously
and then discontinuously with the information transmission
probability. CCA in ER-SF multiplex networks in which two subnetworks
are Erd\H{o}s-R\'{e}nyi (ER) and scale-free (SF) introduces a crossover
from continuous to hybrid growth in adoption size versus information
transmission probability. Our results extend our
understanding of the role of CCA in spreading dynamics, and may elicit
further research.

\end{abstract}

\maketitle

\section{Introduction}

\noindent
Numerous social communication platforms, including Facebook, Twitter,
and email, are a part of the current world-wide information
explosion. These platforms taken together behave as a network of
networks (NON) in which each communication platform functions as a
subnetwork \cite{boccaletti2014structure,gao2012networks,
  d2014networks,kivela2014multilayer}. These NONs can be multilayer,
interdependent, or multiplex. In interdependent networks the
functionality of the components in one network depends on the
functionality of nodes in other networks. A multiplex network is a
special NON case in which each agent can be present in more than one
layer.  Extensive studies of cascading failure, evolutionary games,
synchronization, and spreading dynamics have found that the dynamics of
NONs differ greatly from those of single networks \cite{Kim2013,
  radicchi2014driving,buldyrev2010catastrophic,cozzo2013contact,
  saumell2012epidemic,zhang2015explosive,xu2015suppressed}.  For
example, For example, Buldyrev et al. found a first-order
percolation phase transition in interdependent networks that is qualitatively different from the second-order phase
transition in single networks
\cite{buldyrev2010catastrophic}. Baxter et al. found a hybrid phase transition in the percolation phase transition
on multiplex networks~\cite{baxter2012avalanche}. 

Spreading dynamics in complex networks can be classified as either
biological or social contagions. In biological contagions, such as
epidemic spreading, researchers have found that multilayer networks
promote spreading \cite{cozzo2013contact,saumell2012epidemic}, induce
the coexistence of mixed phase transitions \cite{dickison2012epidemics},
and produce rare-region phenomena \cite{cozzo2016characterization}. In
contrast to biological contagions, social contagions are strongly
affected by social reinforcement \cite{porter2016examples}. Research on
social contagions has focused primarily on a generalized Watts threshold
model \cite{brummitt2012multiplexity,min2014multiple,
  yaugan2012analysis,watts2002simple,lee2014threshold,majdandzic2016multiple}.
As in biological contagions, the behavior of NONs promotes social
contagions \cite{brummitt2012multiplexity}. Majdandzic et al.  found
multiple tipping points of social contagions in multilayer networks
\cite{majdandzic2016multiple, majdandzic2014spontaneous}.  Recently Wang
et al. used a data-driven asymmetric
  socio-biological coevolutionary model to locate an optimal
information diffusion mechanism for suppressing biological contagions
\cite{wang2016suppressing} that enables us to understand the effect of
asymmetry in interacting dynamics
\cite{wang2014asymmetrically,liu2016impacts,granell2013dynamical,funk2009spread}.

Although there are many different communication channels, we usually
select one to transmit information to friends during short periods of
time due to the inelasticity of such resources as time and energy
\cite{holme2012temporal,haerter2012communication, min2016layer}. Thus
when transmitting information through a multiplex network we often
migrate to other channels
\cite{de2014navigability,starnini2016temporal}, and a communication
channel alteration (CCA) occurs. For example, when transmitting
information using texting or email, if we change to another channel a
CCA occurs. This event (i) creates distinct active time periods for
individuals in different communication platforms, and (ii) introduces
time delays when obtaining information from other
platforms. Theoretically CCA induces a non-Markovian effect into the
spreading dynamics that further causes strong dynamic correlations among
the states of neighbors.

Because there has been no systematic study of how CCA affects the
dynamics of social contagions on multiplex networks, we propose a
non-Markovian behavior spreading model on multiplex networks. At any
given time an individual can be active in only one communication layer
and can transmit behavioral information to neighbors and obtain
behavioral information from neighbors only within the same
subnetwork. Using extensive numerical simulations we find that
time delays induced by CCA slow behavior spreading but do not
affect the final adoption size. We also find that CCA suppresses the
final behavior adoption size. It is significant that CCA changes the growth
pattern of the final adoption size on ER-SF networks, i.e., the growth
pattern of the adoption size versus the behavioral information
transmission probability changes from continuous to hybrid.  We
develop a generalized edge-based compartmental method to describe this
non-Markovian spreading model and find that the theoretical predictions
agree with the numerical predictions.

\section{Social contagion model on multiplex networks} \label{sec:model}

\begin{figure}
\begin{center}
\epsfig{file=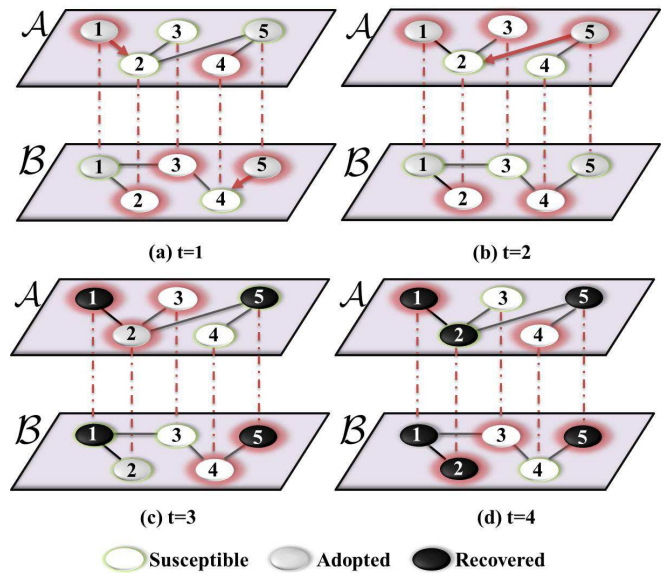,width=1\linewidth}
\caption{(Color online) Illustration of social contagions on multiplex
  networks. Initially, individuals $1$ and $5$ are selected as seeds
  (adopted), and the remaining individuals are susceptible. (a) At time
  $t=1$, individual $1$ ($5$) successfully transmits the information to
  his susceptible neighbor $2$ ($4$) in subnetwork $\mathcal{A}$
  ($\mathcal{B}$). The accumulated pieces of information of $2$ ($4$) in
  subnetwork network $\mathcal{A}$ ($\mathcal{B}$) is
  $m_2^\mathcal{A}=1$ ($m_4^\mathcal{B}=1$). (b) At time $t=2$,
  individuals $1$, $3$ and $5$ are in subnetwork $\mathcal{A}$.
  Individual $5$ successfully transmits the information to $2$, however,
  individual $2$ does not read this information since he belongs to
  subnetwork $\mathcal{B}$. (c) At $t=3$, individual $2$ activates in
  subnetwork $\mathcal{A}$, he fulfills his adoption threshold
  $T_\mathcal{A}=2$. Thus, individual $2$ adopts the
  behavior. Individuals $1$ and $5$ recover with probability
  $\gamma=0.6$. (d) At time $t=4$, all adopted individuals recover and
  no individual whose received information is greater than his adoption
  threshold. The processes terminates.  Each individual activates in
  subnetwork $\mathcal{A}$ with probability $p=0.6$, and in subnetwork
  $\mathcal{B}$ with the complementary probability
  $1-p=0.4$. Individuals activates in the subnetwork are marked with red
  shadow, such as individual $1$ activates in $\mathcal{A}$.}
\label{fig1}
\end{center}
\end{figure}

Communication channels such as
  Facebook, Twitter, and email facilitate interaction and
information-sharing. Taken together they form a multiplex network in
which each communication channel functions as a subnetwork. During short
periods of time individuals select a single communication platform to
transmit information, and then CCA occurs. To understand how CCA affects
the dynamics of social contagion, we examine the behavior spreading
dynamics in two-layer multiplex networks in which each layer or
subnetwork represents a single communication
channel. Figure~\ref{fig1}(a) shows a multiplex network. In the
multiplex network model, subnetworks $\mathcal{A}$ and $\mathcal{B}$
have the same number of nodes, and they randomly match one-to-one, which
means that each individual can communicate with friends through two
different communication channels. To establish the multiplex network we
first assign degrees $k_\mathcal{A}$ and $k_\mathcal{B}$ to individuals
in subnetworks $\mathcal{A}$ and $\mathcal{B}$, respectively, according
to a joint degree distribution $P(\overrightarrow{K}) =P (k_\mathcal{A},
k_\mathcal{B})$. We then build each subnetwork using an uncorrelated
configuration model \cite{Catanzaro2005}.  In the thermodynamic limit,
i.e., the network size $N\rightarrow\infty$, there are no intra-layer
degree correlations in the subnetworks.

We propose a generalized non-Markovian susceptible-adopted-recovered
(SAR) model \cite{wang2015dynamics,wang2016dynamics,wang2015dynamics2}
to describe the behavior spreading dynamics in multiplex networks.
Wang et al.~\cite{wang2015dynamics} found a transition phenomenon in which the dependence of the final adoption size versus information transmission probability can change from discontinuous
to continuous by decreasing the individual adoption threshold, increasing the initial seed size, or enhancing the
network heterogeneity. In
the SAR model, social reinforcement is triggered by the reception of
non-redundant behavioral information, i.e., a susceptible individual
adopts the new behavior only when the amount of received non-redundant
information from neighbors rises above a given adoption threshold.  We
only allow the transmission of non-redundant information.  An individual
in the susceptible state has not adopted the behavior. An individual in
the adopted state has adopted the behavior and is willing to transmit
the information to neighbors through one communication channel. An
individual in the recovered state has lost interest in the behavior and
no longer transmits the information.

During short periods of time the constraints posed by inelastic
resources limit individuals to a single communication channel
\cite{Haerter2012,Holme2012}, i.e., at any given time step an individual
can only be active in one subnetwork. Individuals thus use communication
channels alternation (CCA) as the system
  evolves. To describe CCA, we introduce a layer-switching parameter
$p_i$ and assume that individual $i$ is active in subnetwork
$\mathcal{A}$ with a probability $p_i$ and active in subnetwork
$\mathcal{B}$ with a probability $1-p_i$. For simplicity, we assume that
all individuals have equal values of $p=p_i$. When an individual is
active in a subnetwork $\mathcal{X}\in\{\mathcal{A},\mathcal{B}\}$, it
can transmit the information to neighbors in subnetwork $\mathcal{X}$
and can read information from neighbors in subnetwork $\mathcal{X}$ but
cannot read information in the other non-active subnetwork
$\mathcal{Y}$. Thus individuals cannot simultaneously read all
information from neighbors in all communication channels, and this
introduces \emph{time delays\/} into receiving information.

We use a synchronous updating method \cite{schonfisch1999synchronous} to
renew the states of individuals. We first randomly select a fraction
$\rho_0$ of individuals to be seeds in the adopted state. The remaining
individuals are in the susceptible state.  At each time step, each
adopted individual $v_\mathcal{X}$ active in subnetwork $\mathcal{X}$
transmits the information to each susceptible neighbor $u_\mathcal{X}$
with a probability $\lambda_\mathcal{X}$. When individual
$v_\mathcal{X}$ successfully transmits information to $u_\mathcal{X}$,
the information transmission between them does not occur in subsequent
steps, i.e., we only allow nonredundant information to transmit between
two individuals because each neighbor can only partially guarantee the
credibility and legitimacy of the behavior
\cite{centola2011experimental}.  Note that if a susceptible individual
$u_\mathcal{X}$ is active in subnetwork $\mathcal{X}$, he only reads
information in subnetwork $\mathcal{X}$. Thus the CCA introduces
time delays into receiving information. If an individual
$u_\mathcal{X}$ in subnetwork $\mathcal{X}$ receives a new piece of
information, he adopts the behavior with a probability
$\pi(m_u^\mathcal{A}, m_u^\mathcal{B})$, where the $m_u^\mathcal{X}$
value is the number of cumulative pieces of nonredundant information
from neighbors in subnetwork $\mathcal{X}$. If individual
$u_\mathcal{X}$ adopts the behavior, their counterpart $u_\mathcal{Y}$
also adopts the behavior in subnetwork $\mathcal{Y}$. Here we focus on a
case in which an individual in the susceptible state adopts the new
behavior when $m_u^\mathcal{X}$ exceeds their adoption threshold
$T_u^\mathcal{X}$ in subnetwork $\mathcal{X}$. For simplicity, we assume
all individuals have the same adoption threshold
$T_u^\mathcal{X}=T_\mathcal{X}$ in subnetwork $\mathcal{X}$.  The
$\pi(m_u^\mathcal{A},m_u^\mathcal{B})$ value is
\begin{equation}\label{pi_The}
\pi(m_u^\mathcal{A},m_u^\mathcal{B})=
\left\{
\begin{array}{rcl}
1, ~m_u^\mathcal{A}\geq T_\mathcal{A}~ \mathrm{or}~ m_u^\mathcal{B}\geq
T_\mathcal{B},\\
0, ~\mathrm{others}.
\end{array} \right.
\end{equation}
To include the social reinforcement effect, both $T_\mathcal{A}$ and
$T_\mathcal{B}$ values are greater than unity. At each time step we
assume that all individuals in the adopted state lose interest in
transmitting the information and with a probability $\gamma$ they
recover. The spreading stops when all the adopted individuals become
recovered, and the received information of all susceptible
individuals does not exceed their threshold in either subnetwork.
Figure~\ref{fig1} shows the behavior spreading dynamics in multiplex
networks.

There are two key features in our proposed
spreading dynamics. (1) The memory effect is induced in our model. Unlike bootstrap percolation~\cite{baxter2010bootstrap,baxter2011heterogeneous,
di2018insights} or a
threshold model~\cite{watts2002simple} in which a susceptible individual becomes active (or adopted) only when its current number
or fraction of adopted neighbors is larger than a given value, in our model a susceptible individual becomes
adopted when their received accumulated information in either subnetwork is larger than a threshold. (2) CCA is
included in our model. Unlike the models in Refs.~\cite{brummitt2012multiplexity,
yaugan2012analysis} in which each node can obtain the information from
two subnetworks simultaneously, CCA allows an individual to be active in only one subnetwork at a time step.

\section{Theoretical method}

\noindent
From the description of the behavior adoption process in
Sec.~\ref{sec:model} we know that there is a \emph{non-Markovian\/}
characteristic in the dynamics because (i) social reinforcement occurs
when non-redundant behavioral information transmissions are remembered,
and (ii) CCA causes time delays in the reception of information. This
non-Markovian characteristic makes the strong dynamic correlations among
the states of the neighbors difficult to describe. Here we develop a
generalized edge-based compartmental method
\cite{wang2014epidemic,miller2013incorporating,miller2011note,
  wang2016unification} of describing the spreading of behavior in
multiplex networks. In this theoretical method we
  assume that the networks are large, the edges sparse, there are no
  degree-degree correlations, and the dynamics evolve continuously.

An individual $u$ adopting the new behavior must take into account their
received information in both subnetworks $\mathcal{A}$ and
$\mathcal{B}$. Denoting $u_\mathcal{X}$ as an individual $u$ active in
subnetwork $\mathcal{X}\in\{\mathcal{A},\mathcal{B}\}$, we quantify the
probability that individual $u$ is in the susceptible state and assume
that $u_\mathcal{X}$ is in the cavity state \cite{Karrer2010}, i.e.,
that $u_\mathcal{X}$ cannot transmit the information to neighbors in
subnetwork $\mathcal{X}$ but can receive information from neighbors in
all the subnetworks. The probability that an individual $v_\mathcal{X}$
has not transmitted the information to a neighbor $u_\mathcal{X}$ along
a randomly chosen edge in subnetwork $\mathcal{X}$ by time $t$ is
$\theta_\mathcal{X}(t)$. At time $t$, the probability that
$u_\mathcal{X}$ has received units of information $m_\mathcal{X}$ from
subnetwork $\mathcal{X}$ is
\begin{equation}\label{phi_A}
\phi_{m_\mathcal{X}}^\mathcal{X}(k_\mathcal{X},t)=
{k_\mathcal{X} \choose m_\mathcal{X}}
[\theta_\mathcal{X}(t)]^{k_\mathcal{X}
-m_\mathcal{X}}[1-\theta_\mathcal{X}(t)]^{m_\mathcal{X}},
\end{equation}
where $k_\mathcal{X}$ is the degree of individual $u$ in subnetwork
$\mathcal{X}$. Here the CCA disallows individual $u_\mathcal{X}$ from
reading all information from all subnetworks because he can only focus
on the information in the layer in which he is currently active. At
time $t$, individual $u$ can only read received information from
neighbors in subnetwork $\mathcal{X}$ when he is active in that
subnetwork, i.e., when $u$ is active in $\mathcal{X}$ he cannot read
received information in $\mathcal{Y}$.

\begin{figure}
\begin{center}
\epsfig{file=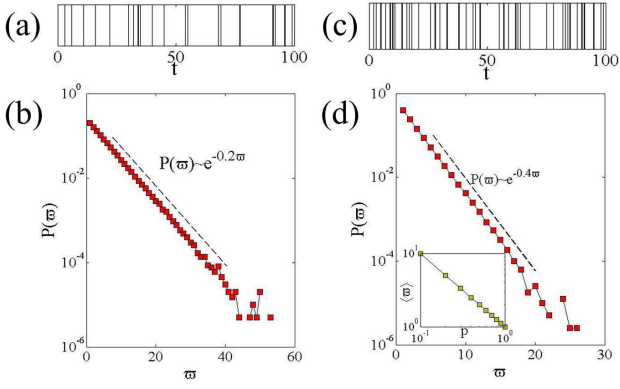,width=1\linewidth}
\caption{(Color online) Dynamics of social contagions on RR-RR multiplex
  networks. Active events in subnetwork $\mathcal{A}$, i.e., individuals
  are active in subnetwork $\mathcal{A}$, versus time $\varpi$ with (a)
  $p=0.8$ and (c) $p=0.6$, respectively. The time interval distribution
  $P(\varpi)$ between two consecutive activates in subnetwork
  $\mathcal{A}$ with (b) $p=0.8$ and (d) $p=0.6$, respectively. We set
  other parameters to be $\rho_0=0.1$, $\gamma=1.0$, $k=10$,
  $\lambda_\mathcal{A}=\lambda_\mathcal{B}=0.9$, and $T_\mathcal{A}=
  T_\mathcal{A}=3$.}
\label{illu}
\end{center}
\end{figure}

The CCA between two subnetworks for individual $u$ is a stochastic
Poisson process.  The time interval distribution between two successive
actions of $u$ in subnetwork $\mathcal{A}$ is $P (\overline{\omega})\sim
e ^{ -(1-p) \overline{\omega}}$ \cite{Mieghem2014}, as shown in
Figs.~\ref{illu}(a)--(b) and (d)--(e) where $p$ is the probability that
$u$ is active in subnetwork $\mathcal{A}$. Note that we use random regular (RR)
  \cite{newman2010networks} networks to describe the two subnetworks
  $\mathcal{A}$ and $\mathcal{B}$ in Fig.~\ref{illu}. All nodes have the
  same degree in the RR network, i.e., $P_\mathcal{A}(k_\mathcal{A})
  =P_\mathcal{B}(k_\mathcal{B})=1$ if $\langle k_\mathcal{A}
  \rangle=\langle k_\mathcal{B}\rangle=10$. Each RR network can be built
  using an uncorrelated configuration model \cite{Catanzaro2005}.  We
find the approximate average time interval of all individuals active in
subnetwork $\mathcal{X}$ to be
\begin{equation}\label{interval}
\langle\overline{\omega}_\mathcal{X}\rangle= \frac{1}{1-p_\mathcal{X}},
\end{equation}
where $p_\mathcal{X}$ is the probability that an individual is active in
subnetwork $\mathcal{X}$. When $\mathcal{X}=\mathcal{A}$, then
$p_\mathcal{X}=p$. When $\mathcal{X}=\mathcal{B}$, then
$p_\mathcal{X}=1-p$.

If individual $u$
is active in subnetwork $\mathcal{Y}$ at time $t$, on average his
latest active time in subnetwork $\mathcal{X}$ is
$t-\langle\overline{\omega}_\mathcal{X}\rangle$. Here he reads all
information in subnetwork $\mathcal{Y}$.  The approximate number of
information units he can read in subnetwork $\mathcal{X}$ is the number
of information units in the inbox of subnetwork $\mathcal{X}$ at
$t-\langle\overline{\omega}_\mathcal{X}\rangle$. The cumulative number
of information units $n_\mathcal{X}$ read by susceptible individual
$u_\mathcal{X}$ at time $t$ has a probability
\begin{equation}\label{phi_A_delay}
\chi_{n_\mathcal{X}}^\mathcal{X}(k_\mathcal{X},t) =
{k_\mathcal{X} \choose n_\mathcal{X}}
[\theta_\mathcal{X}(t-\langle\overline{\omega}_\mathcal{X}\rangle)]^{k_\mathcal{X}-n_\mathcal{X}}
 [1-\theta_\mathcal{X}(t
-\langle\overline{\omega}_\mathcal{X}\rangle)]^{n_\mathcal{X}}.
\end{equation}

If individual $u$ with degree $\overrightarrow{K}=(k_\mathcal{A},
k_\mathcal{B})$ has not adopted the behavior by time $t$, the cumulative
pieces of information individual $u$ has read in both subnetworks
$\mathcal{A}$ and $\mathcal{B}$ will be less than the adoption
thresholds $T_\mathcal{A}$ and $T_\mathcal{B}$, respectively. At time
$t$, if individual $u$ is active in subnetwork $\mathcal{A}$ he remains
susceptible with a probability
\begin{equation}\label{F_A}
\begin{split}
F_\mathcal{A}(\overrightarrow{K},
t) &=\sum_{m_\mathcal{A}=0}^{k_\mathcal{A}}\sum_{n_\mathcal{B}=0}^{k_\mathcal{B}}
\phi_{m_\mathcal{A}}^\mathcal{A}(k_\mathcal{A},
t)\chi_{n_\mathcal{B}}^\mathcal{B}(k_\mathcal{B},t) \\
&\times\prod_{j_\mathcal{A}=0}^{m_\mathcal{A}}\prod_{j_\mathcal{B}=0}^{n_\mathcal{B}}
[1-\pi(j_\mathcal{\mathcal{A}}, j_\mathcal{B})].
\end{split}
\end{equation}
Similarly, if individual $u$ is active in subnetwork $\mathcal{B}$ at time
$t$ he remains susceptible with a probability
\begin{equation}\label{F_B}
\begin{split}
F_\mathcal{B}(\overrightarrow{K},
t) &=\sum_{n_\mathcal{A}=0}^{k_\mathcal{A}}\sum_{m_\mathcal{B}=0}^{k_\mathcal{B}}
\chi_{n_\mathcal{A}}^A(k_\mathcal{A},t)
\phi_{m_\mathcal{B}}^\mathcal{B}(k_\mathcal{B},t) \\ &\times\prod_{j_\mathcal{A}=0}^{n_\mathcal{A}}
\prod_{j_\mathcal{B}=0}^{m_\mathcal{B}}
[1-\pi(j_\mathcal{A}, j_\mathcal{B})].
\end{split}
\end{equation}
The probability that individual $u$ with a degree $\overrightarrow{K}$
is susceptible is
\begin{equation}\label{S_K}
s(\overrightarrow{K},t)=(1-\rho_0)[p_\mathcal{A}F_\mathcal{A}
(\overrightarrow{K},
t)+p_\mathcal{B}F_\mathcal{B}(\overrightarrow{K}, t)],
\end{equation}
where the factor $(1-\rho_0)$ is the probability that individual $u$ is
initially susceptible, and $p_\mathcal{X}=\frac{1}{N}\sum_{u=1}^N
p_u^\mathcal{X}$ is the average probability that individual $u$ is
active in subnetwork $\mathcal{X}$. Note that $p_\mathcal{A}
=1-p_\mathcal{B}$. Examining the degree distribution
$P(\overrightarrow{K})$, the fraction of susceptible individuals at time
$t$ is
\begin{equation}\label{S}
S(t)=\sum_{\overrightarrow{K}} P
(\overrightarrow{K})s(\overrightarrow{K},t).
\end{equation}

According to the definition of $\theta_\mathcal{X}$, an endpoint
individual $v_\mathcal{X}$ of the randomly selected edge of
$u_\mathcal{X}$ can be in a susceptible, adopted, or recovered state,
thus $\theta_\mathcal{X}(t)$ can be rewritten
\begin{equation}\label{theta_A}
\theta_\mathcal{X}(t)=\xi_S^\mathcal{X}(t)+
\xi_A^\mathcal{X}(t)+\xi_R^\mathcal{X}(t),
\end{equation}
where $\xi_S^\mathcal{X}(t)$ [$\xi_A^\mathcal{X}(t)$ or
$\xi_R^\mathcal{X}(t)$] is the probability that individual
$v_\mathcal{X}$ is susceptible (adopted or recovered) and has not
transmitted the information to $u_\mathcal{X}$ by time $t$.

If individual $u_\mathcal{X}$ is initially susceptible, he is in the
cavity state and thus cannot transmit the information to susceptible
neighbors $v_\mathcal{X}$ in subnetworks $\mathcal{X}$. When individual
$v_\mathcal{X}$ with degree $k_\mathcal{X}^\prime$ is susceptible and
active in subnetwork $\mathcal{X}$, he can only receive the information
from $k_\mathcal{X}^\prime-1$ other neighbors in subnetwork $\mathcal{X}$. At time $t$, if the
susceptible individual $v_\mathcal{X}$ is active in subnetwork $\mathcal{X}$ the
probability that he can read $m_\mathcal{X}$ pieces of information
from this subnetwork is 
\begin{equation}\label{phi_A_Nei}
\tau_{m_\mathcal{X}}^\mathcal{X}(k_\mathcal{X}^\prime,t)=
{k_\mathcal{X}^\prime-1 \choose m_\mathcal{X}} \theta_\mathcal{X}
(t)^{k_\mathcal{X}^\prime-m_\mathcal{X}-1}
[1-\theta_\mathcal{X}(t)]^{m_\mathcal{X}}.
\end{equation}
In contrast, individual $v$ can only read the information at time
$t-\langle\overline{\omega}_\mathcal{X}\rangle$ in subnetwork
$\mathcal{X}$ when he is active in subnetwork $\mathcal{Y}$ at time
$t$. As in Eq.~(\ref{phi_A_delay}), the probability that individual
$v_\mathcal{X}$ reads $n_\mathcal{X}$ pieces of information is
\begin{equation}\label{phi_A_Nei_delay}
\begin{split}
\varsigma_{n_\mathcal{X}}^\mathcal{X}(k_\mathcal{X}^\prime,t) &=
{k_\mathcal{X}^\prime-1 \choose n_\mathcal{X}} \theta_\mathcal{X}
(t-\langle\overline{\omega}_\mathcal{X}\rangle)^{k_\mathcal{X}^\prime-n_\mathcal{X}-1}
\\
&\times[1-\theta_\mathcal{X}
(t-\langle\overline{\omega}_\mathcal{X}\rangle)]^{n_\mathcal{X}}.
\end{split}
\end{equation}
Individual $v_\mathcal{X}$ remains susceptible when the cumulative
pieces of information read from neighbors in subnetwork $\mathcal{X}$ is
lower than adoption threshold $T_\mathcal{X}$ in the absence of
individual $u_\mathcal{X}$. When $v$ is active in subnetwork
$\mathcal{X}$, he reads $n_\mathcal{Y}$ pieces of information in
subnetwork $\mathcal{Y}$ with a probability
$\chi_{n_\mathcal{Y}}^\mathcal{Y}(k_\mathcal{Y}^\prime,t)$ [see
  Eq.~(\ref{phi_A_delay})]. Thus the probability that individual $v$ is
susceptible is
\begin{equation}\label{S_k_A_Nei_acitvie_A}
\begin{split}
\Psi_\mathcal{X}^\mathcal{X}(\overrightarrow{K}, t) &=
\sum_{m_\mathcal{X}^\prime=0}^{k_\mathcal{X}^\prime-1}
\sum_{n_\mathcal{Y}^\prime=0}^{k_\mathcal{Y}^\prime}
\tau_{m_\mathcal{X}^\prime}^\mathcal{X}(k_\mathcal{X}
^\prime,t)\chi_{n_\mathcal{Y}^\prime}^\mathcal{Y}
(k_\mathcal{Y}^\prime,t)\\
&\times
\prod_{j_\mathcal{X}^\prime=0}^{m_\mathcal{X}^\prime}
\prod_{j_\mathcal{Y}^\prime=0}^{n_\mathcal{Y}^\prime}
[1-\pi(j_\mathcal{X}^\prime,j_\mathcal{Y}^\prime)].
\end{split}
\end{equation}
Similarly, when individual $v$ is active in subnetwork $\mathcal{Y}$
the probability that $v$ will read $n_\mathcal{X}$ and $m_\mathcal{Y}$
pieces of information from subnetworks $\mathcal{X}$ and $\mathcal{Y}$
is $\varsigma_{n_\mathcal{X}}^\mathcal{X}(k_\mathcal{X}^\prime,t)$ and
$\phi_{m_\mathcal{Y}}^\mathcal{Y}(k_\mathcal{Y}^\prime,t)$,
respectively. The probability that individual $v$ is susceptible is
\begin{equation}\label{S_k_A_Nei_acitvie_B}
\begin{split}
\Psi_\mathcal{X}^\mathcal{Y}(\overrightarrow{K}, t)& =
\sum_{n_\mathcal{X}^\prime=0}^{k_\mathcal{X}^\prime-1}
\sum_{m_\mathcal{Y}^\prime=0}^{k_\mathcal{Y}^\prime}
\varsigma_{n_\mathcal{X}^\prime}^\mathcal{X}(k_\mathcal{X}^\prime,t)
\phi_{m_\mathcal{Y}^\prime}^\mathcal{Y}(k_\mathcal{Y}^\prime,t) \\
&\times\prod_{j_\mathcal{X}^\prime=0}^{n_\mathcal{X}^\prime}
\prod_{j_\mathcal{Y}^\prime=0}^{m_\mathcal{Y}^\prime}
[1-\pi(j_\mathcal{X}^\prime,j_\mathcal{Y}^\prime)].
\end{split}
\end{equation}
When an initially susceptible individual $v$ is active in subnetwork
$\mathcal{X}$, the probability that he remains susceptible is
\begin{equation}\label{S_k_A_Nei}
\Theta_\mathcal{X}(\overrightarrow{K},t)=
p_\mathcal{X}\Psi_\mathcal{X}^\mathcal{X}(\overrightarrow{K},
t)+p_\mathcal{Y}\Psi_\mathcal{X}^\mathcal{Y}(\overrightarrow{K},t).
\end{equation}
Denoting
  $H_\mathcal{X}(k_\mathcal{X}^\prime |k_\mathcal{X})$ to be the
  probability of a node with degree $k_\mathcal{X}$ connects to a node
  with degree $k_\mathcal{X}^\prime$ in network $\mathcal{X}$.  Thus the
  probability that individual $u_\mathcal{X}$ connects to a susceptible
  individual in subnetwork $\mathcal{X}$ is
\begin{equation}\label{xi_S_A_n}
\xi_S^\mathcal{X}(t)=(1-\rho_0)
\sum_{\overrightarrow{K}}H_\mathcal{X}(k_\mathcal{X}^\prime
|k_\mathcal{X})\Theta_\mathcal{X}(\overrightarrow{K},t).
\end{equation}
In an uncorrelated network, $H_\mathcal{X}(k_\mathcal{X}^\prime
|k_\mathcal{X}) =k_\mathcal{X}^\prime P(\overrightarrow{K})/\langle
k_\mathcal{X}\rangle$.  We rewrite Eq.~(\ref{xi_S_A_n}) to be
\begin{equation}\label{xi_S_A}
\xi_S^\mathcal{X}(t)=(1-\rho_0)\frac{1}{\langle
k_\mathcal{X}\rangle}\sum_{\overrightarrow{K}}k_\mathcal{X}^\prime P
(\overrightarrow{K})\Theta_\mathcal{X}(\overrightarrow{K},t).
\end{equation}
When the information transmits through an edge at time $t$ in subnetwork
$\mathcal{X}$, the edge does not fulfill the definition of
$\theta_\mathcal{X}(t)$. The decreasing of $\theta_\mathcal{X}(t)$ is
thus
\begin{equation}\label{D_theta_A}
\frac{d\theta_\mathcal{X}(t)}{dt}
=-p_\mathcal{X}\lambda_\mathcal{X}\xi_A^\mathcal{X}(t).
\end{equation}
For $\xi_R^\mathcal{X}(t)$ to grow, (i) the information cannot be
transmitted through the edge, and (ii) the adopted individual must
recover at time $t$. The evolution of $\xi_R^\mathcal{X}(t)$
is
\begin{equation}\label{xi_R_A_2}
\frac{d\xi_R^\mathcal{X}(t)}{dt}=\gamma(1-p_\mathcal{X}
\lambda_\mathcal{X})\xi_A^\mathcal{X}(t).
\end{equation}
Combining Eqs.~(\ref{D_theta_A}) and (\ref{xi_R_A_2}) and the initial
condition $\theta_\mathcal{X}(0)=1$ and $\xi_R^\mathcal{X}(0)=0$ gives
us
\begin{equation}\label{xi_R_A}
\xi_R^\mathcal{X}(t)=\frac{\gamma(1-p_\mathcal{X}\lambda_\mathcal{X})
[1-\theta_\mathcal{X}(t)]}{p_\mathcal{X}\lambda_\mathcal{X}}.
\end{equation}
We use Eqs.~(\ref{theta_A}), (\ref{xi_S_A}), ($\ref{D_theta_A}$), and
(\ref{xi_R_A}) to obtain the value of $\theta_\mathcal{X}(t)$.

Using the evolution process of the behavior spreading dynamics described
in Sec.~\ref{sec:model}, we derive the evolution equations of the
fraction of individuals in the adopted and recovered states,
\begin{equation}\label{A_t}
\frac{dA(t)}{dt}=-\frac{dS(t)}{dt}-\gamma A(t),
\end{equation}
and
\begin{equation}\label{R_t}
\frac{dR(t)}{dt}=\gamma A(t),
\end{equation}
respectively. Combining Eqs.~(\ref{S}) and (\ref{A_t})--(\ref{R_t}), we
obtain the time evolution of the behavior spreading dynamics in
multiplex networks.  When $t\rightarrow\infty$, the final behavior
adoption size is denoted $R(\infty)$.

We next examine the growth pattern of the final behavior adoption size
$R(\infty)$ versus the information transmission probability
$\overrightarrow{\lambda}=(\lambda_\mathcal{A},\lambda_\mathcal{B})$ and
the CCA probability $p$. We first investigate the effects of the
time delays induced by CCA on the dynamics of social contagions by
comparing them with a null model without time delays.  Details about the
null model are supplied in the Appendix.  Figure~\ref{illu2}(a) shows
that the time delays induced CCA affect behavior spreading dynamics,
including the time evolutions of susceptible $S(t)$, adopted $A(t)$, and
recovered $R(t)$ individuals. We find that $S(t)$ [$R(t)$] decreases
(increases) with $t$, and that $A(t)$ first increases and then
decreases. Note that the time delays induced by CCA slow the behavior
adoption. To quantify the slowing caused by the time delays induced by
CCA, we compute the stabilizing time of the system $t_{\rm max}$, i.e.,
the average time needed for the system to reach the final state. When
$0\leq p\leq 0.5$, $t_{\rm max}$ first increases with $p$ and then
decreases. When the $p$ value is small, most individuals are active in
subnetwork $\mathcal{B}$. When susceptible individuals receive
information that exceeds their adoption threshold, and they quickly
adopt the behavior. When the $p$ value is increased, some individuals
become active in subnetwork $\mathcal{A}$ but adopt the behavior only
after their received information exceeds their adoption threshold. Thus
$t_{\rm max}$ first increases. When the $p$ value is large, fewer
individuals adopt the behavior (see Fig.~\ref{fig2}) and $t_{\rm max}$
thus decreases.  The inset of Fig.~\ref{illu2}(b) shows the retardation
time $\Delta t_{\rm max}$ for social contagion models with and without
time delays. Note that $\Delta t_{\rm max}$ first increases with $p$ and
then decreases. We find the same phenomena for $0.5\leq p\leq1.0$
because subnetworks $\mathcal{A}$ and $\mathcal{B}$ are both RR
networks.

Figure~\ref{illu2}(a) shows that the time delays induced by CCA do not
affect the final behavior adoption size $R(\infty)$. This is because
when a susceptible individual fulfills the behavior adoption conditions
the time delays only affect the behavior adoption time. This also
indicates that the critical points of the system remain the same when
there are no time delays in their behavior adoption, i.e., a susceptible
individual adopts the new behavior as soon as the received pieces of
information equal or exceed the adoption threshold, independent of the
subnetwork in which he is currently active. Thus when examining the
final state of the behavior we disregard time delays, i.e.,
Eqs.~(\ref{phi_A_delay}) and (\ref{phi_A_Nei_delay}) are the same as
Eqs.~(\ref{phi_A}) and (\ref{phi_A_Nei}), respectively. We perform
numerical simulations and theoretical analyses of the social contagions
on ER-SF multiplex networks, and find that the time delays induced by
CCA slow the spreading dynamics but do not affect the final adoption
size.

\begin{figure}
\begin{center}
\epsfig{file=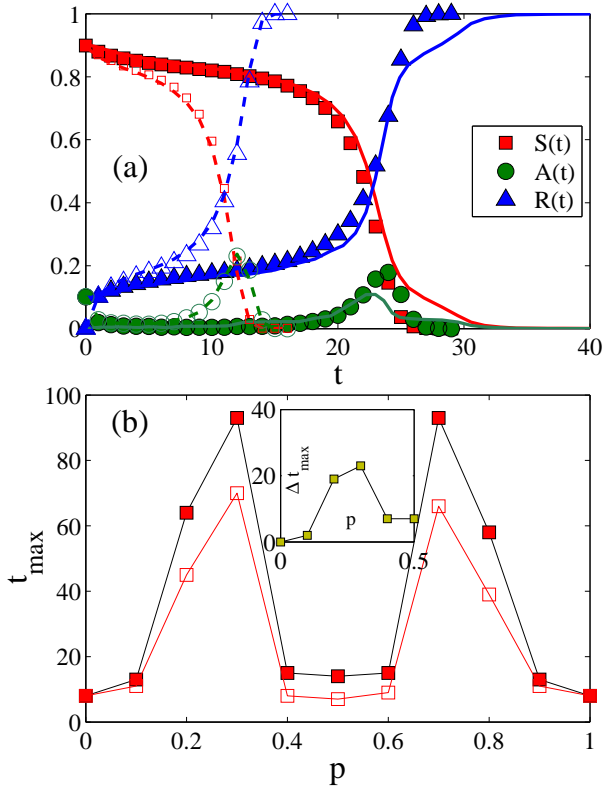,width=1\linewidth}
\caption{(Color online) Dynamics of social contagions on RR-RR multiplex
  networks. (a) The time evolutions of susceptible $S(t)$, adopted
  $A(t)$ and recovered $R(t)$ individuals with (solid symbols) and
  without (empty symbols) time delays when $p=0.8$.  The symbols are
  numerical simulation results and lines are the theoretical
  predictions. (b) The stabilizing time of the system $t_{\rm max}$ with
  (solid symbols) and without (empty symbols) time delays versus
  $p$. The inset of (b) shows the retardation time as a function of $p$
  for the system with or without time delays.  We set other parameters
  to be $\rho_0=0.1$, $\gamma=1.0$, $k=10$,
  $\lambda_\mathcal{A}=\lambda_\mathcal{B}=0.9$, and $T_\mathcal{A}=
  T_\mathcal{A}=3$.}
\label{illu2}
\end{center}
\end{figure}

In the final state, i.e., when $t\rightarrow\infty$,
there are no nodes in the adopted state,
  and no information is transmitted through edges. Thus we have
$d\theta_\mathcal{X}(t)/dt=0$ and
$\theta_\mathcal{X}(t)=\theta_\mathcal{X}
(t-\langle\overline{\omega}_\mathcal{X}\rangle)=\theta_
\mathcal{X}^\ast$. Combining Eqs.~(\ref{theta_A}),
(\ref{xi_S_A})--(\ref{D_theta_A}), and (\ref{xi_R_A}) we obtain
\begin{equation}\label{D_theta_A_Stady}
\begin{split}
\theta_\mathcal{X}^\ast   &=(1-\rho_0)\frac{\sum_{\overrightarrow{K}}k_\mathcal{X}
P(\overrightarrow{K})\Theta_\mathcal{X}(\overrightarrow{K},\infty)}{\langle
k_\mathcal{X}\rangle}
\\ &+\frac{\gamma(1-p_\mathcal{X}\lambda_\mathcal{X})[1-\theta_\mathcal{X}^\ast]}
{p_\mathcal{X}\lambda_\mathcal{X}} \\
 &=f_\mathcal{X}
(\theta_\mathcal{A}^\ast,\theta_\mathcal{B}^\ast).
\end{split}
\end{equation}

When $\rho_0\rightarrow 0$, then $\theta_\mathcal{X}^*=1$ is a trivial
solution of Eq.~(\ref{D_theta_A_Stady}), but this vanishingly small
fraction of seeds cannot trigger global behavior adoption because
$T_\mathcal{X}>1$ \cite{wang2015dynamics}. To stimulate global behavior
adoption, we must have a finite fraction of seed individuals. Here
$\theta_\mathcal{X}^*=1$ is no longer the solution of Eq.~(\ref{D_theta_A_Stady}), which
now has either one or three fixed points (including multiplicity).
If Eq.~(\ref{D_theta_A_Stady}) has only one solution at all values of $\lambda_\mathcal{X}$,
then $\theta_\mathcal{X}^*$ decreases continuously with
$\lambda_\mathcal{X}$, and this leads to a continuous growth pattern in
the final behavior adoption $R(\infty)$. If the number of the solutions
of Eq.~(\ref{D_theta_A_Stady}) varies with $\lambda_\mathcal{X}$, the situation is
different. For a given $\lambda_\mathcal{X}$, if there is only one fixed
point of Eq.~(\ref{D_theta_A_Stady}), it is the physically meaningful solution. If there
are three fixed points, which are stable, unstable, and saddle points,
only the maximum solution is physically meaningful in our irreversible
behavior spreading dynamics when we randomly select a relatively small
fraction of seeds, since the individuals in the adopted state
persistently transmit the information to their neighbors, and
$\theta_\mathcal{X}^*$ decreases from unity. Thus a saddle-node
bifurcation occurs~\cite{strogatz2014nonlinear,
gleeson2007seed}.
Through a bifurcation analysis of
Eq.~(\ref{D_theta_A_Stady}),
we find that the system undergoes a cusp catastrophe: varying
$\lambda_\mathcal{X}$ the physically meaningful stable solution of
$\theta_\mathcal{X}^*$ suddenly produces a different outcome. 
Therefore, the
growth patten of $R(\infty)$ will be discontinuous because a meaningful
solution decreases abruptly at such critical conditions as the critical
information transmission probability $\overrightarrow{\lambda}_c$ and
the critical CCA probability $p_c$.

To determine $p_c$, we first rewrite Eq.~(\ref{D_theta_A_Stady}) to be
\begin{equation}\label{add}
F_\mathcal{A}(\theta_\mathcal{A}^\ast,\theta_\mathcal{B}^\ast)=\theta_\mathcal{A}^\ast
-f_\mathcal{A}(\theta_\mathcal{A}^\ast,\theta_\mathcal{B}^\ast)=0,
\end{equation}
and
\begin{equation}\label{add2}
F_\mathcal{B}(\theta_\mathcal{A}^\ast,\theta_\mathcal{B}^\ast)=\theta_\mathcal{B}^\ast
-f_\mathcal{B}(\theta_\mathcal{A}^\ast,\theta_\mathcal{B}^\ast)=0.
\end{equation}
At the discontinuous critical point, the curves of $f_\mathcal{A}
(\theta_\mathcal{A}^\ast, \theta_\mathcal{B}^\ast)$ and
$f_\mathcal{B}(\theta_\mathcal{A}^\ast, \theta_\mathcal{B}^\ast)$ are
tangent to each other at the discontinuous critical point
\cite{yuan2017eradicating,parshani2010interdependent}. Thus we find
that the critical point for the
  discontinuous growth pattern is given by
$\frac{d\theta_\mathcal{A}^\ast(\theta_\mathcal{B}^\ast)}
{d\theta_\mathcal{B}^\ast}\frac{d\theta_\mathcal{B}^\ast(\theta_\mathcal{A}^\ast)}
{d\theta_\mathcal{A}^\ast}=1$. Combining Eqs.~(\ref{add}) and
(\ref{add2}), we further obtain the discontinuous critical points of
$\overrightarrow{\lambda}_c$ and $p_c$ by solving both
Eq.~(\ref{D_theta_A_Stady}) and
\begin{equation}\label{Critical_Conditation}
\frac{\partial f_\mathcal{A}(\theta_\mathcal{A}^\ast,\theta_\mathcal{B}^\ast)}
{\partial\theta_\mathcal{B}^\ast}
\frac{\partial f_\mathcal{B}(\theta_\mathcal{A}^\ast,\theta_\mathcal{B}^\ast)}
{\partial\theta_\mathcal{A}^\ast}=1.
\end{equation}

\section{Numerical simulations}

\noindent
Here we perform extensive simulations on multiplex networks, including
RR-RR networks [i.e., in which each subnetwork is a random regular (RR)
  network] and ER-SF networks [i.e., in which subnetworks $\mathcal{A}$
  and $\mathcal{B}$ are Erd\H{o}s-R\'{e}nyi (ER) \cite{Erdos1959} and
  scale-free (SF) \cite{Catanzaro2005} networks, respectively]. In each
case we set the network size, average degree, and recovery probability
to be $N=10^4$, $\rho_0=0.1$, $\langle k_\mathcal{A}\rangle=\langle
k_\mathcal{B} \rangle=10$, and $\gamma=1.0$, respectively, unless stated
otherwise.

\subsection{RR-RR multiplex networks}

\begin{figure}
\begin{center}
\epsfig{file=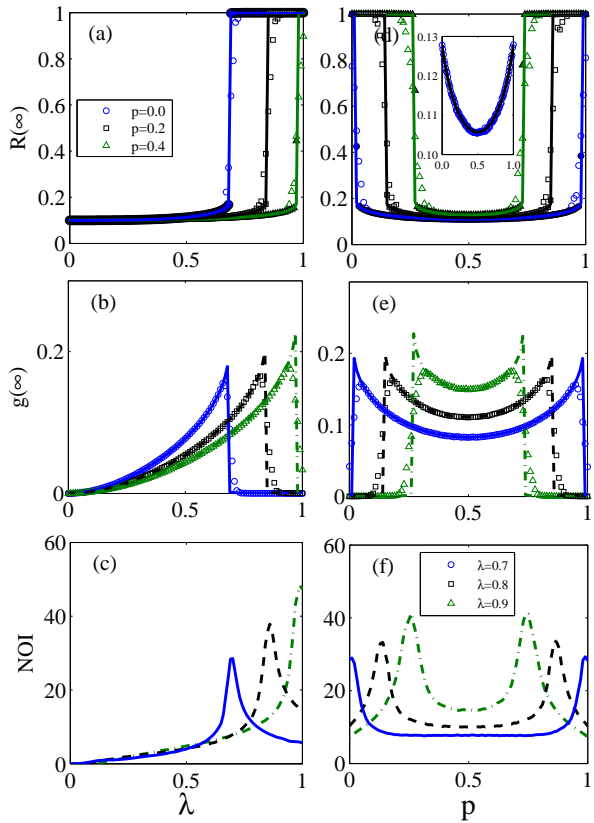,width=1\linewidth}
\caption{(Color online) Behavior spreading on RR-RR networks.  (a) Final
  behavior adoption size $R(\infty)$, (b) the fraction of individuals in
  the subcritical state $g(\infty)$, and (c) NOI versus information
  transmission probability $\lambda$ for CCA probability $p=0.0$, $0.2$
  and $0.4$. (d) $R(\infty)$, (e) $g(\infty)$, and (f) NOI versus $p$
  for $\lambda=0.7$, $0.8$ and $0.9$.  The inset of (d) shows
  $R(\infty)$ versus $p$ for $\lambda=0.6$. The empty (full) symbols are
  the simulated values of $R(\infty)$ at network size $N=10^4$
  ($N=10^6$). The lines in (a), (b), (d) and (e) are the theoretical
  values, and are numerical simulation results in (c) and (f).  Other
  parameters are $\rho_0=0.1$, $\gamma=1.0$, $\langle k\rangle=10$, and
  $T_\mathcal{A}=T_\mathcal{A}=3$, respectively.}
\label{fig2}
\end{center}
\end{figure}

\noindent
We first study social contagions on RR-RR multiplex networks in which
each node in each subnetwork has a degree $k=10$. Figure~\ref{fig2}(a)
shows $R(\infty)$ as a function of $\lambda$ under different CCA
probabilities $p$. We find that $R(\infty)$ first
grows continuously for small values of $\lambda$, and then increases
{\it discontinuously\/} at $\lambda_c$, i.e., exhibiting a hybrid
growth, regardless of $p$. We can understand the
discontinuous increasing of $R(\infty)$ by studying the fraction of
individuals $g(\infty)$ in the subcritical state.  When an individual is
in the subcritical state he has received the information but have not
adopted the behavior, and the number of information units from distinct
neighbors is one less than the adoption threshold in subnetwork
$\mathcal{A}$ or $\mathcal{B}$ \cite{wang2015dynamics}.  Slightly
increasing the value of $\lambda$ can increase the number of subcritical
state individuals with information units equal to or greater than their
threshold [see Fig.~\ref{fig2}(b)] and lead to a discontinuous jump in
$R(\infty)$. In addition, $R(\infty)$ decreases with $p$, since an
increasing number of susceptible individuals are unable to fulfill the
behavior adoption condition. Because near the critical point the system exhibits a critical slowing,
we locate the numerical critical point $\lambda_c^{\rm I}$ by examining
the number of iterations (NOI), which is widely used in the bootstrap
percolation~[41,42] and cascading failures~\cite{baxter2010bootstrap,
baxter2011heterogeneous,di2018insights} and cascading failures~\cite{gao2011robustness,liu2016breakdown}, required to reach the
final state, and we take into account only iterations in which at least
one individual adopts the behavior. Figure~\ref{fig2}(c)
shows that at the critical point the NOI exhibit a peak. Note that our
theoretical method accurately predicts $R(\infty)$ and the growth
pattern of $R(\infty)$. The deviations around the critical points are
caused by finite-size network effects [see Fig.~\ref{fig2}(a)].

Figures~\ref{fig2}(d)--\ref{fig2}(f) show the effect of $p$ on social
contagions. Unlike when
  $T_\mathcal{A}=T_\mathcal{B} =1$ in Fig.~\ref{r_add} of Appendix B,
  Fig.~\ref{fig2}(d) shows that the final behavior adoption size
  $R(\infty)$ versus $p$ is non-monotonic. Specifically, for relatively
small values of $\lambda=0.6$ few susceptible individuals adopt the
behavior, and $R(\infty)$ first decreases {\it continuously\/} with $p$
and then increases {\it continuously\/} [see the inset in
  Fig.~\ref{fig2}(d)].  For large values of $\lambda=0.7,0.8,0.9$,
$R(\infty)$ first decreases \emph{discontinuously\/} with $p$ and then
increases \emph{discontinuously\/} [see Fig.~\ref{fig2}(d)].  We can
similarly understand the growth pattern of $R(\infty)$ by studying
$g(\infty)$. For a small value of $p$, e.g., $p=0.1$ when $\lambda=0.9$,
most individuals in subnetwork $\mathcal{A}$ are active, many
individuals adopt the behavior, and few individuals remain in the
subcritical state [see Fig.~\ref{fig2}(e)].  With an increase of $p$,
fewer individuals are active in subnetwork $\mathcal{B}$, many
individuals in subnetwork $\mathcal{A}$ are in the subcritical state
[see Fig.~\ref{fig2}(e)], and there is a sharp decrease in
$R(\infty)$. By further increasing $p$, many individuals ``jump''
between subnetworks $\mathcal{A}$ and $\mathcal{B}$, and fewer
individuals receive one fewer information units than the adoption
threshold in the two subnetworks. Thus $g(\infty)$ decreases [see
  Fig.~\ref{fig2}(e)]. Similarly $g(\infty)$ first increases and then
decreases discontinuously when the value of $p$ is large. Note that NOI
versus $p$ exhibits two peaks at $p_c^1=1-p_c^2$ because the two
subnetworks are RR networks [see Fig.~\ref{fig2}(f)].

\begin{figure}
\begin{center}
\epsfig{file=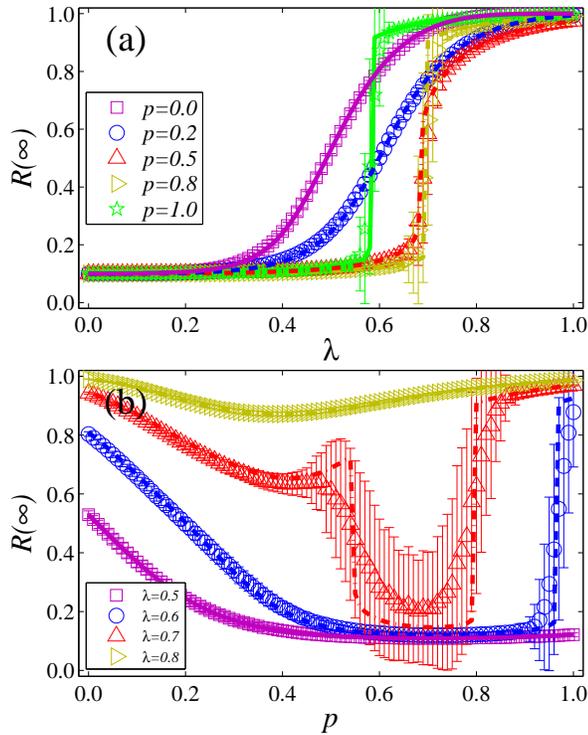,width=1\linewidth}
\caption{(Color online) Dynamics of social contagions on uncorrelated
  ER-SF multiplex networks.  (a) The final behavior adoption size
  $R(\infty)$ versus information transmission probability $\lambda$
  under different communication channels alternation probability
  $p$. (b) $R(\infty)$ as a function of $p$ for different
  $\lambda$. Lines are the theoretical predictions from Eqs.~(\ref{S_K})
  and (\ref{D_theta_A})-(\ref{xi_R_A}). The error bars indicate the
  standard deviations. Other parameters are set to be $\langle
  k_\mathcal{A}\rangle=10$, $\langle k_\mathcal{B}\rangle=10$,
  $\upsilon_\mathcal{B}=3.0$, $\rho_0=0.1$, and $T_\mathcal{A}=
  T_\mathcal{A}=3$.}
\label{fig3}
\end{center}
\end{figure}

\subsection{ER-SF multiplex networks}

\noindent
When studying social contagions on ER-SF multiplex networks, we assume
that there are no degree-degree correlations in the intralayers and
interlayers.  We generate the SF networks using the same method as that
used in uncorrelated configuration networks that have a power-low degree
distribution $P_\mathcal{B}(k_\mathcal{B})
=\frac{1}{\sum_{k_\mathcal{B}}
k_\mathcal{B}^{-v_\mathcal{B}}}k_\mathcal{B}^{-v_\mathcal{B}}$.
In network $\mathcal{B}$ without
  degree-degree correlations, the maximum degree follows a structural
  cut-off \cite{boguna2004cut}, i.e., $k_{\rm max}\sim \sqrt{N}$. The SF
  network is built using the uncorrelated configuration method in
  Ref.~\cite{Catanzaro2005}. Figure~\ref{fig3} shows the social
contagions on ER-SF networks. Figure~\ref{fig3}(a) shows that CCA
changes the growth pattern of $R(\infty)$ versus $\lambda$. When $p=0.0$
($1.0$), the number of individuals only active in subnetwork
$\mathcal{B}$ ($\mathcal{A}$) and $R(\infty)$ grows continuously
(discontinuously) versus $\lambda$ (see
Ref.~\cite{wang2015dynamics}). Increasing $p$ increases the number of
individuals active in the homogeneous subnetwork $\mathcal{A}$, and
there are more individuals in the subcritical state who are likely to
simultaneously adopt the behavior. Thus we see a hybrid growth
in $R(\infty)$, i.e., $R(\infty)$
first grows continuously for small values of $\lambda$, and then
grows discontinuously at $\lambda_c$. 

\begin{figure}
\begin{center}
\epsfig{file=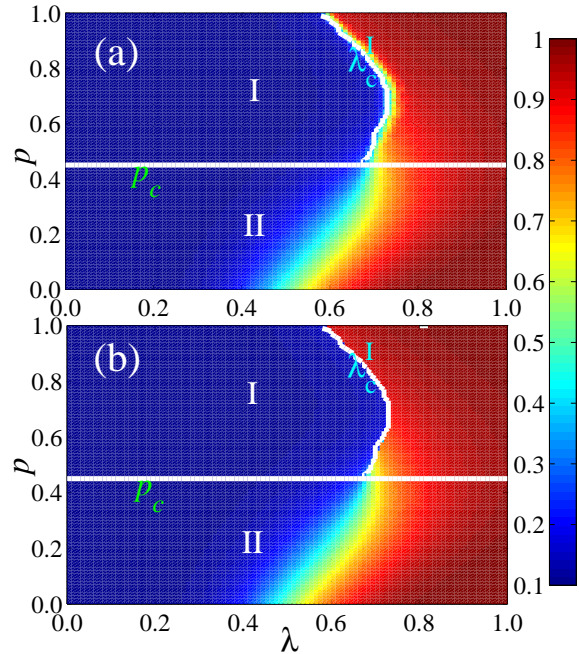,width=0.9\linewidth}
\caption{(Color online) Final behavior adoption size $R(\infty)$ versus
  information transmission probability $\lambda$ and communication
  channels alternation probability $p$ on uncorrelated ER-SF
  networks. Color-coded values of $R(\infty)$ from numerical simulations
  (a) and theoretical solutions (b) in the parameter plane
  $(\lambda,p)$. The white horizontal line $p_c$ separates the plane
  into regions I and II. In region I (II), $R(\infty)$ exhibits
  a hybrid (continuous) growth with $\lambda$. The white curves denote
  the theoretical discontinuous critical points $\lambda_c^{\rm I}$.  We
  set other parameters as $\langle k_\mathcal{A}\rangle=10$, $\langle
  k_\mathcal{B}\rangle=10$, $\upsilon_\mathcal{B}=3.0$, $\rho_0=0.1$,
  and $T_\mathcal{A}=T_\mathcal{A}=3$.}
\label{fig4}
\end{center}
\end{figure}

Figure~\ref{fig3}(b) shows that $R(\infty)$ versus $p$ exhibits
differing patterns under different values of $\lambda$. For a small
value $\lambda=0.5$, $R(\infty)$ first decreases continuously with $p$.
For a relatively large value, e.g., $\lambda=0.6$, $R(\infty)$ first
decreases continuously with $p$ and then increases discontinuously. Note
that when $\lambda=0.7$, $R(\infty)$ first decreases continuously with
$p$, then increases to a peak at some $p$, then decreases
discontinuously, and finally increases discontinuously. For a very large
value $\lambda=0.9$, $R(\infty)$ first decreases continuously and then
increases continuously. For intermediate values of $0<p<1$, the CCA
emerges and constrains user ability to receive enough information to
exceed the adoption threshold of a subnetwork, and thus there is a
non-monotonous varying of $R(\infty)$. We can understand the different
growth patterns by studying high degree nodes (hubs), which (i) are more
likely to adopt the behavior than those in homogeneous network networks
for small values of $\lambda$, and (ii) lead to individuals adopting the
behavior gradually with fewer individuals in the subcritical state
simultaneously adopting the behavior \cite{wang2015dynamics}. The
fraction of hubs in the live subnetwork varies with $p$. We define a
live subnetwork to be a user's active subnetwork at different time
steps. For small $\lambda=0.5$, individuals active in heterogeneous
(homogeneous) networks can easily (with difficulty) adopt the behavior,
adding the second role of hubs, and $R(\infty)$ thus decreases with
$p$. For the larger $\lambda=0.6$, $R(\infty)$ first decreases
continuously because some individuals are active in heterogeneous
networks when $p$ is relatively small (e.g., $p<0.5$), and $R(\infty)$
then increases discontinuously because many individuals are active in
homogeneous networks and in the subcritical state for large $p$. The
strange phenomena when $\lambda=0.7$ is also induced by the live
network. When $p\approx 0.4$, many individuals are active in homogeneous
subnetwork $\mathcal{A}$, and $R(\infty)$ thus increases with $p$. A
further increase of $p$ is a slight perturbation that moves many
individuals into the subcritical state, and $R(\infty)$ decreases
discontinuously.  When $\lambda=0.8$, most individuals adopt the
behavior, few individuals are in the subcritical state, and there is
only a non-monotonous varying of $R(\infty)$ versus $p$.
These phenomena do not occur when
  $T_\mathcal{A}=T_\mathcal{B}=1$, as shown in Fig.~\ref{r_add}. We
can predict these phenomena using our theoretical method.  Note that the
deviations near the discontinuous points can be eliminated by enlarging
the network size $N$.

Figure~\ref{fig4} shows $R(\infty)$ versus the $\lambda-p$ plane. Using
the growth pattern of $R(\infty)$ versus $\lambda$, we divide the plane
into regions I and II according to a critical CCA probability $p_c$.  In
region I, $R(\infty)$ first grows continuously with $\lambda$ then increases discontinuously at $\lambda_c$, and in
region II, $R(\infty)$ grows continuously with $\lambda$. There is thus
a crossover phenomenon in the growth pattern: When $p>p_c$, the growth
pattern of $R(\infty)$ is discontinuous; otherwise it is continuous. We
can explain the growth pattern of $R(\infty)$ using bifurcation
theory. The discontinuous critical points $\lambda_c^{\rm I}$ exhibit a
non-monotonic change with $p$ because of the CCA.  When $p>p_c$,
individuals are more likely to be active in subnetwork $\mathcal{A}$,
and thus there are some hubs and many low-degree individuals in the live
subnetwork. These hubs promote behavior adoption. Increasing $p$
decreases the number of hubs and low-degree individuals in the live
network, and $p_c$ thus first increases then decreases.  The results
from our theoretical method agree with those from the numerical
simulations.

\section{Conclusions}

\noindent
We have investigated how communication channel alteration (CCA) affects
the dynamics of social contagions. We first propose a non-Markovian
behavior spreading model for multiplex networks in which each individual
can only transmit and obtain the information from neighbors in their own
subnetwork. To include CCA, we assume that an individual can be active
in only one communication layer and at any given time can only transmit
behavioral information to neighbors and read behavioral information from
neighbors within the same subnetwork.  The CCA slows a user's ability to
receive information from both subnetworks. Thus time delays in obtaining
the information from both subnetworks are introduced.  We then perform
numerical simulations of artificial multiplex networks and find that the
time delays caused by CCA slow the behavior adoption process but do not
affect the final behavior adoption size. In addition, CCA suppresses the
final behavior adoption size $R(\infty)$ but does not change the growth
pattern of $R(\infty)$ on RR-RR networks. We find in ER-SF networks that
the growth pattern of $R(\infty)$ can be changed from hybrid to
continuous by decreasing the layer-switching probability. To quantify
the non-Markovian spreading dynamics, we develop an edge-based
compartmental method that produces results that agree with the numerical
simulation results.

We have examined how CCA---an important inter-layer switching
mechanism---affects social contagions. We first construct the
connections between human dynamics
\cite{barabasi2005origin,vazquez2005exact, brockmann2006scaling} and
social contagions on multiplex networks.  Individuals accomplishing
different tasks using different communication channels do so in patterns
that exhibits memory and burst characteristics. Our results here are the
first to investigate the effects of human dynamics on social contagions
in multiplex networks, and they expand our understanding of phase
transition phenomena. The hybrid growth in the final adoption size is similar to the hybrid phase
transition observed in other dynamics~\cite{dorogovtsev2008critical,di2018multiple,wu2014multiple}, and the critical phenomena of our proposed social contagions near
the critical point need further investigation Our theoretical method allows us to understand how
CCA shapes spreading dynamics and to analyze different dynamic processes
on multiplex networks. Our work may stimulate further research on social
contagions that takes into account both realistic spreading mechanisms
and network topologies and provides theoretical insights into how to
control the spread of epidemics. In addition, social contagions with a
heterogeneous layer-switching probability is an intriguing subject for
examination \cite{starnini2016temporal}. For example, a layer-switching
probability follows a power-law distribution, and the layer-switching
probability of each individual is dependent on their inherent
characteristics.

\section*{Acknowledgements}

\noindent
This work was partially supported by the National Natural Science
Foundation of China (Grant Nos.~11575041, and 61672238), China
Postdoctoral Science Foundation (CPSF) (Grant No. 2018M631073), the
Fundamental Research Funds for the Central Universities, and Natural
Science Foundation of Shanghai (Grant No. 18ZR1412200). L.A.B. acknowledges
the support of UNMdP, PICT 0429/13, and  DTRA Grant HDTRA1-14-1-0017. The Boston University Center for Polymer Studies is supported by NSF Grants No. PHY-1505000, No. CMMI-1125290, and No. 624 CHE-1213217, and by
DOE Contract No. DE-AC07-05Id14517.

\section*{Appendix A: Null model for social contagions with CCA}

\noindent
We use the layer-switching parameter $p$ to describe CCA in the null
model. The only difference between this null model and the model
described in Sec.~2 is that in this null model we assume a susceptible
individual becomes adopted when the accumulated units of received
information is equal to or larger than the adoption threshold in any
subnetwork, regardless of whether the individual is active in the
subnetwork. This difference allows the susceptible individual to obtain
the behavioral information without time delays. The null model includes
CCA but not time delays.

\section*{Appendix B: $T_\mathcal{A}=T_\mathcal{B}=1$}

\begin{figure}
\begin{center}
\epsfig{file=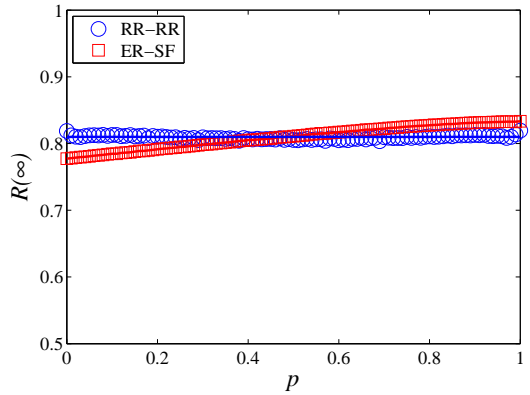,width=0.9\linewidth}
\caption{(Color online) Final behavior
    adoption size $R(\infty)$ versus communication channels alternation
    probability $p$ on multiplex networks when
    $T_\mathcal{A}=T_\mathcal{B}=1$. The symbols are the numerical
    simulation results, and lines are the theoretical predictions. We
    set other parameters as $\langle k_\mathcal{A}\rangle=10$,
    $\rho_0=0.1$, $\langle k_\mathcal{B}\rangle=10$ and
    $\upsilon_\mathcal{B}=3.0$ for SF networks.}
\label{r_add}
\end{center}
\end{figure}

 Figure~\ref{r_add} shows the contagions
  on multiplex networks with $T_\mathcal{A}=T_\mathcal{B}=1$, i.e., the
  contagions on both subnetworks return to the simple contagion. We find
  that $R(\infty)$ does not change with $p$ on RR-RR multiplex networks
  and increases with $p$ on ER-SF networks.

\end{document}